\documentclass{article}
\usepackage{spconf,hyperref}
\usepackage{textcase}
\usepackage{epsfig,setspace,amsmath,epsf,amssymb,bm,theorem,cite,graphicx, epstopdf,algorithm,float,color,mathtools}
\usepackage[table,xcdraw]{xcolor}
\usepackage{dsfont}

\newcolumntype{P}[1]{>{\centering\arraybackslash}p{#1}}
\newcolumntype{M}[1]{>{\centering\arraybackslash}m{#1}}

\newtheorem{theorem}{Theorem}

\newenvironment{Proof}[1]{\medskip\par\noindent{\bf Proof:\,}\,#1}{{\mbox{\,$\blacksquare$}\par}}

\newcommand{\R}{\mathbb{R}}

\newcommand{\eps}{\varepsilon}
\renewcommand{\epsilon}{\eps}
\renewcommand{\P}{\mathbb{P}}

\newcommand{\E}{\mathbb{E}}

\newcommand{\cN}{\mathcal{N}}

\newcommand{\Z}{\mathbb{Z}}

\allowdisplaybreaks

\title{Deriving Moments in the Age of Gossip Process from Percolation}

\name{Thomas Jacob~Maranzatto and Sennur Ulukus}

\address{\normalsize{Department of Electrical and Computer Engineering, 
University of Maryland, College Park, MD}}

\begin{document}
\maketitle

\begin{abstract}
    This paper concerns fundamental identities in the study of age of information (AoI) in gossip networks.  We recover known recursive identities for arbitrary $k$th moments of the age process based on the recent connection between AoI and first passage percolation.  Apart from the connection to percolation, our proofs are more concise and can be followed using only elementary facts from probability.  Our argument generalizes some techniques known in the statistical physics community, and we remark on connections to the Eden model.
\end{abstract}


\section{Introduction}
Given a network of interacting users, a fundamental question is to quantify the freshness of their stored data.  This paper considers the \textit{age of information} (AoI) metric over systems of gossiping users, where users send packets to neighbors randomly.  The AoI metric was first introduced by Kaul, Yates, and Gruteser~\cite{aoiseminal}, and we direct the reader to the survey~\cite{Yates_survey} for background on related metrics over a wide class of communication and queuing models.

Our focus is on a gossiping model introduced by Yates~\cite{yates_gossip_spawc}.  Here, a directed network of $N$ users $G = (\cN, E)$ communicate in a decentralized manner by sending packets to their immediate neighbors. Without loss of generality, take $\cN = [0,\ldots, N]$, and distinguish the node 0 (sometimes referred to as node $n_0$) as the source so only directed edges of the type $(0,u)$ exist, never $(u,0)$.  The source acts as the \emph{state of the world}, and does not control information flow or packets sent in the gossip network.  If $(u,v) \in E$, then $u$ sends packets to $v$ via a Poisson process with rate $\lambda_{uv}$, independent of all other processes in the network. For every time $t \ge 0$, each node $u$ stores a real number $N_u(t)$ denoting the timestamp of its current packet.  We initialize $N_u(0) = 0$ for all $u$, and for all $t \ge 0$ the source maintains $N_{0}(t) = t$.  When the Poisson clock $(u,v)$ rings at time $t$, the counter of $v$ updates as $N_v(t) = \max\{N_v(t), N_u(t)\}$. The \textit{age of information} (AoI) of $v$ is defined as $X_v(t) := t - N_v(t)$.  It is helpful to think of the AoI as the time elapsed since the packet stored at node $v$ was generated by the source.

This model and variants have been studied extensively~\cite{Buyukates_2022, gossip_dynamic,   kaswan2023age, Kaswan_2023, maranzatto2024age, gossip_grid,  yates_gossip_isit, yates_gossip_spawc}, and we point the reader to a survey for more context~\cite{kaswan2023age}.  Majority of the work in the area has used the stochastic hybrid system (SHS) framework, and the most well-known example of this is Yates' recursive formula for the long-term expected age of a subset of vertices~\cite{yates_gossip_isit,yates_gossip_spawc} which we briefly recall.  Define the total rate of $u$ into a subset of vertices $S$ as 
\begin{align}
    \lambda_u(S) = \begin{cases}
    \sum_{v \in S} \lambda_{uv}, &u\not\in S,\\
    0,&u \in S,
    \end{cases}
\end{align}
and let $v_S = \lim_{t \to \infty} \E[ \min_{u\in S}N_u(t)]$ be the long-term expected minimum age of any vertex in $S$.  Then, for any subset $S$, Yates showed,
\begin{align}\label{eq:yates}
    v_S = \frac{1 + \sum_{u\not\in S} \lambda_u(S)v_{S \cup\{u\}}}{\lambda_0(S) +\sum_{u\not\in S} \lambda_u(S)}.
\end{align}

\begin{figure}
    \centering
    \includegraphics[width=0.6\linewidth]{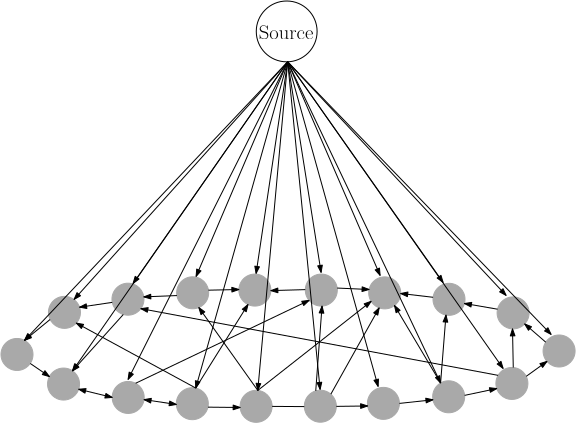}
    \caption{A typical gossip network we consider.  In general, the source does not have to update all vertices with the same rate.}
    \label{fig:gossip}
\end{figure}

Similar recursive identities for the $k$th moments were recovered by Abd-Elmagid and Dhillon~\cite{AbdElmagid2023DistributionOT}.  Letting $v_S^k = \lim_{t \to \infty} \E \left[(\min_{u\in S}N_u(t))^k\right]$ be the long-term average $k$th moment of the age process, they proved,
\begin{align}\label{eq:abd}
    v_S^k = \frac{kv_S^{k-1} + \sum_{u\not\in S} \lambda_u(S)v^k_{S \cup\{u\}}}{\lambda_0(S) +\sum_{u\not\in S} \lambda_u(S)}.
\end{align}

Our contribution is to reprove these fundamental identities without the use of the SHS framework. The authors of~\cite{AbdElmagid2023DistributionOT} proved many other identities for the moment generating function of the age process, and generalized to joint moment identities.  We comment on how our approach applies to the joint moment recursion in the conclusion. We also remark that \eqref{eq:yates} was generalized in~\cite[eqn.~(5)]{yates_gossip_isit} to a source which self updates with rate $\lambda_e$;  this is perhaps the more well-known recursive formulation.  Our analysis focuses on the standard AoI formulation instead of the \textit{version} AoI, however our technique readily extends to this case.

\subsection{Prerequisites and Contributions}
While the existing proofs of \eqref{eq:yates} and \eqref{eq:abd} are fairly short, they require the heavy machinery in the form of the SHS framework.  Our contribution is a short proof of both identities from the \emph{first principles}.  Our proof relies on the recent connection between AoI in Poissonian gossip networks and the first passage percolation~\cite{maranzatto2024fpp}.  Let us first recall the definition of the first passage percolation.  Let $\gamma$ be any path in an edge-weighted directed graph $G = (\cN, E, w)$, where the weights may be random variables.  Then, define $T_G(\gamma) = \sum_{e \in \gamma}w_e$ as the \textit{passage time} of $\gamma$.  The \textit{first passage time} from $u$ to $v$ is defined as $T_G(u,v) = \min_{\gamma}T(\gamma)$, where the minimum is over paths with $u$ as the starting node and $v$ as the final node.  For any two subsets of vertices $S, S'$, the passage time from $S$ to $S'$ is naturally $T_G(S,S') = \min_{u \in S, v \in S'} T_G(u,v)$.  Note that if $G$ is finite and the weights $w_e$ admit a density, then there is a unique path $\gamma'$ achieving $T_G(S,S')$ almost surely.

Given a gossip network $G = (\cN, E)$ with Poisson clocks and update rates $\{\lambda_e\}_{e \in E}$, the authors of~\cite{maranzatto2024fpp} construct an auxiliary weighted graph $G' = (\cN, E, w)$ on the same vertex and edge set, but with edge weights defined as $w_{uv} = \operatorname{Exp}(\lambda_{uv})$, i.e., edge weights are exponential random variables with parameter corresponding to the associated Poisson process rate.

It is important to note that while $G'$ has random edge weights there is no stochastic process associated to each edge, therefore, the first passage time between vertices can be computed.  We make use of the following theorem, which relates the first passage time in $G'$ to the AoI of a vertex in $G$.   

\begin{figure}
    \centering
    \includegraphics[width=0.5\linewidth]{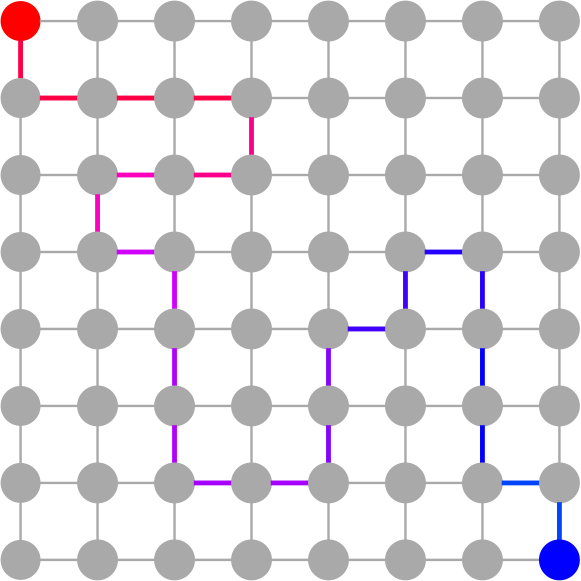}
    \caption{A sample first passage path on a grid graph from the red vertex to the blue vertex.  One can view the first passage time as the time a packet would take to travel from the red vertex to the blue vertex, along a minimizing random path.}
    \label{fig:placeholder}
\end{figure}

\begin{theorem}[~\cite{maranzatto2024fpp}] \label{thm:fpp}
    For any locally finite gossip network $G = (\cN, E)$ and $S\subset \cN$, a.s. there exists some time $t_0 < \infty$ such that for any $t > t_0$ the age of information of $S$ has distribution given by,
    \begin{equation} 
    X_S(t) = T_{G'}(n_0,S). 
    \end{equation}
\end{theorem}

Theorem~\ref{thm:fpp} was originally expressed in terms of a slightly different graph $G'$ with reversed edges, and $t_0$ was excluded in favor of a more precise characterization of the age process.\footnote{The time $t_0$ can be chosen explicitly by the following $t_0 = \inf \{t: u \text{ receives a timestamped packet from the source at time } t \}$.}    This was necessary for the proof, but the weaker formulation above is sufficient for our purposes and we believe makes our presentation cleaner.

In their original form, \eqref{eq:yates} and \eqref{eq:abd} only hold in the limit, therefore, our proofs are a modest strengthening of these results as they hold for some large but finite (random) time.  Moreover, the new proofs trivially extend to locally finite graphs\footnote{Locally finite graphs have countably many vertices, and each vertex has finite neighborhood.  An example of a locally finite graph is $\Z^d$, the $d$-dimensional integer lattice.}, where it is unclear apriori if the SHS framework can be extended to countably infinite state transitions.  Finally, we remark that applying the SHS framework requires stability of the underlying process.  Stability was implicitly assumed in the above works, thus, this paper can be seen as a verification of this assumption.  As an example of why this is important, in attempts to generalize the SHS framework to dynamic networks, we have found cases where stability does not hold.

\subsection{Connection to the Eden Model}
Our proof is inspired by techniques pioneered in statistical physics for the Eden model.  Classical questions in the Eden model are closely related to those recently raised about age of gossip.  While the specifics of the Eden model and Eden conjecture are outside the scope of this paper, there are very similar recursions in the seminal work of Dhar~\cite{Dhar1988FirstPP,Dhar1986AsymptoticSO} to those of Yates~\cite{yates_gossip_spawc}.  Dhar's recursions were generalized by Bertrand and Pertinand~\cite{Bertrand_2018}, but still remained upper bounds on the first moment of the first passage time.  Upper bounds on the second moment were studied recently by Auffinger and Tang~\cite{Auffinger_revisited}, which was a direct inspiration for our proof for the $k$th moment.  While the proofs are similar, the works mentioned above recovered upper bound inequalities, and we are able to push slightly further to equality.  We point the interested reader to the survey~\cite{auffinger2017fifty} for more open questions in percolation and the Eden model.

\section{Moments from First Principles}
\begin{figure}
    \centering
    \includegraphics[width=0.5\linewidth]{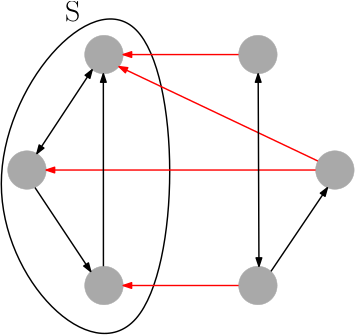}
    \caption{In the proofs of Theorems~\ref{thm:first_moment} and~\ref{thm:moment}, the random variable $Y$ is one of those edges in $E(S)$, highlighted in red above.  The black edges are not included in $E(S)$.  We have not included the source vertex $n_0$, but implicitly, the edges from $n_0$ to the network are still present, though not in $E(S)$.}
    \label{fig:subset_recur}
\end{figure}

For any locally finite gossip network $G$, let $X_S := X_S(t)$ be the distribution of the age of information of subset $S$; this distribution is time-invariant for large enough $t$, and for the rest of the paper, we assume that $t$ is large enough for this invariance to hold.  By the discussion above, one can equivalently view $X_S$ as the first passage time from $n_0$ to $S$ in the dual graph $G'$, so that $X_S \stackrel{D}{=} T_{G'}(n_0, S)$.  In what follows, for any set of vertices $S$, we use $E(S)$ to denote the in-edge boundary of $S$, minus those edges adjacent to $n_0$; since $G$ is directed, these directed edges originate outside $S$ and terminate in $S$; see Fig.~\ref{fig:subset_recur} for an example.  We implicitly assume $n_0 \not\in S$.  For any edge $e = (u,v)$, we will use the notation $S \cup e$ to mean $S \cup \{u,v\}$.  Using this notation, the presentation of the main theorems are slightly different from~\eqref{eq:yates},~\eqref{eq:abd}, but will make our algebraic manipulations more convenient.  To see the equivalence, notice that for any function $f : 2^\cN \to \R$, we can re-index the sum and combine like terms to obtain,
\begin{align}
    \sum_{e \in E(S)} \lambda_e f(S \cup e) = \sum_{i \in N(S)} \lambda_i(S) f(S \cup i).
\end{align}
Applying this change of variable recovers the original form found in~\cite{yates_gossip_isit, yates_gossip_spawc, AbdElmagid2023DistributionOT}.

We choose to present our results in two parts:  Theorem~\ref{thm:first_moment} gives a new proof for Yates' recursion of the first moment, and Theorem~\ref{thm:moment} gives the higher moments in~\cite{AbdElmagid2023DistributionOT}.  Our reasoning is that, Yates' recursion is much more prominent in the literature, and we feel the standalone proof is pedagogically useful, though Theorem~\ref{thm:moment} encompasses all cases and is our main contribution in this paper.  We also believe this first moment proof is useful as it avoids much of the tedious algebra in the sequel.  The following argument is inspired by the classical works of Dhar~\cite{Dhar1988FirstPP, Dhar1986AsymptoticSO} on the Eden model.

\begin{theorem}[\cite{yates_gossip_spawc}, Theorem~1]\label{thm:first_moment}
    For any subset $S \subset G$ and large enough time $t$, the AoI of $S$ satisfies,
    \begin{align}
    \E[X_S] = \frac{1 + \sum\limits_{e \in E(S)} \lambda_e \E[X_{S\cup e}]}{{\lambda_0(S) + \sum\limits_{e \in E(S)}\lambda_e}}.
    \end{align}
\end{theorem}

\begin{Proof}
    Choose $t >t_0$ as in Theorem~\ref{thm:fpp}.  For any subset $S$, let $Y$ be the random variable indexing the next edge that rings in $E(S)$; the notation $\{Y = e\}$ then denotes the event that edge $e$ is the first edge to ring among all boundary edges of $S$.  Let $D = \lambda_0(S) + \sum\limits_{e \in E(S)}\lambda_e$ denote the total update rate into the set $S$.  We need the following fact, which is a restatement of the well-known formula for the index of the minimum of exponential random variables,  
    \begin{align}\label{eq:prob_index}
        \P[Y=e] = \frac{\lambda_e}{D}.
    \end{align}
    Furthermore, by the Markov property and the equivalence $X_S = T_{G'}(n_0, S)$ as distributions, we have,
    \begin{align} \label{eq:first_cond_exp}
        \E[X_S | Y=e ] = \begin{cases}
            \E[\operatorname{Exp}(D)],  &n_0 \in e,\\
             \E[X_{S\cup e} + \operatorname{Exp}(D)], &n_0 \not \in e.
            \end{cases}
    \end{align}
    By the law of total expectation, we have,
    \begin{align}
      \E[X_S] =& \E[\E[X_S|Y]]\\
      =& \sum_{e \in E(S)} \E[X_S | Y=e]\P[Y=e]\\
      =&  \frac{1}{D}\Biggr(\sum_{\substack{
          e\in E(S)\\ n_0 \in e}
      }\lambda_e \E[\operatorname{Exp}(D)] \nonumber\\
      &  + \sum_{\substack{
          e\in E(S)\\ n_0 \not\in e}
      }\lambda_e \E[X_{S\cup e} + \operatorname{Exp}(D)] \Biggr) \label{SU1}\\
      =& \frac{1}{D}\Biggr(\frac{\lambda_0(S)}{D} + \sum_{\substack{
          e\in E(S)\\ n_0 \not\in e}
      }\lambda_e\E[X_{S\cup e}]  + \frac{\lambda_e}{D}\Biggr) \label{SU2}\\
      =& \frac{1}{D}\Biggr(1 + \sum_{\substack{
          e\in E(S)\\ n_0 \not\in e}
      }\lambda_e \E X_{S\cup e} \Biggr), \label{SU3}
   \end{align}
   where \eqref{SU1} is obtained by applying \eqref{eq:first_cond_exp}, and \eqref{SU2} is obtained by expanding the expectations. Then, \eqref{SU3} follows by combining terms as $\lambda_0/D + \sum\lambda_e/D = 1$.
\end{Proof}

The proof technique for an arbitrary $k$th moment is essentially the same.  We will again apply the law of total expectation conditioned with respect to $Y$.  The difference is there are now terms of the form $\E[(X + \operatorname{Exp}(D))^k]$ that we have to deal with.  By independence, we can expand this expectation using the binomial theorem and the rest of the proof is rearranging and combining the resulting terms.

\begin{theorem}[\cite{AbdElmagid2023DistributionOT} Theorem 1]\label{thm:moment}
    For any subset $S \subset G$ and large enough time $t$, the $k$th AoI moment of $S$ satisfies,
    \begin{align}
        \E[X_S^k] = \frac{k\E[X_S^{k-1}] + \sum\limits_{e \in E(S)} \lambda_e \E[X_{S\cup e}^k] }{{\lambda_0(S) + \sum\limits_{e \in E(S)}\lambda_e}}.
    \end{align}
\end{theorem}

\begin{Proof}
    Define $t, Y, D$ as above. Before starting, we need the following generalization of \eqref{eq:first_cond_exp},
    \begin{align}\label{eq:expectation_ids}
        \E[X_S^k | Y=e ] = \begin{cases}
            \E[\operatorname{Exp}(D)^k],  &n_0 \in e,\\
            \E[(X_{S\cup e} + \operatorname{Exp}(D))^k], &n_0 \not \in e,
        \end{cases}
    \end{align}
    which follows by considering the conditional distribution of $X_S^k$ given $\{Y = e\}$ for the two classes of $e$.  Furthermore, the exponential is independent from $X_{S\cup e}$ by the Markov property.  Again, by the law of total expectation,
    \begin{align}
        \E[X_S^k] =& \E[\E[X_S^k | Y]]\\
        =& \sum_{e \in E(S)} \E[X_S^k | Y=e]\P[Y=e]\\
        =& \frac{1}{D}\Biggr(\sum_{\substack{
              e\in E(S)\\ n_0 \in e}
          }\lambda_e \E[\operatorname{Exp}(D)^k] \nonumber \\
        &  + \sum_{\substack{
              e\in E(S)\\ n_0 \not\in e}
        }\lambda_e \E[(X_{S\cup e} + \operatorname{Exp}(D))^k] \Biggr).
    \end{align}
    Since $X_{S\cup e}$ and the exponential are independent, we can expand the right-hand term, as
    \begin{align}
        & \frac{1}{D}\Biggr(\sum_{\substack{
              e\in E(S)\\ n_0 \in e}
          }\lambda_e\E[\operatorname{Exp}(D)^k] \nonumber\\
        &\quad + \sum_{\substack{
              e\in E(S)\\ n_0 \not\in e}
          }\lambda_e \sum_{j=0}^{k} {k \choose j} \E[\operatorname{Exp}(D)^j]\E[X_{S\cup e}^{k-j}]\Biggr) \nonumber\\  
        &=  \frac{1}{D}\Biggr(\lambda_0(S)\E[\operatorname{Exp}(D)^k] \nonumber\\
            & \quad  + \sum_{\substack{e\in E(S)\\ n_0 \not\in e}}\lambda_e \sum_{j=1}^{k} {k \choose j} \E[\operatorname{Exp}(D)^j]\E[X_{S\cup e}^{k-j}]\Biggr) \nonumber\\
            &\quad+\frac{1}{D}\Biggr(\sum_{\substack{e\in E(S)\\ n_0 \not\in e}} \lambda_e \E[X_{S\cup e}^k]\Biggr), \label{eq:split}
    \end{align}
    where in the second step we combined the $n_0 \in e$ terms and split up the inner sum according to the $j=0$ term.  We now demonstrate the first term in~\eqref{eq:split} is actually $\frac{k}{D} \E[X_S^{k-1}]$, which will complete the proof.  To that end, we write,
    \begin{align}
          & \lambda_0(S)\E[\operatorname{Exp}(D)^k] \nonumber\\
          &\quad +\sum_{\substack{
              e\in E(S)\\ n_0 \not\in e}
          }\lambda_e \sum_{j=1}^{k} {k \choose j} \E[\operatorname{Exp}(D)^j]\E[X_{S\cup e}^{k-j}] \nonumber \\
          &= \frac{k}{D}\lambda_0(S)\E[\operatorname{Exp}(D)^{k-1}] \nonumber \\
          &\quad+ \sum_{\substack{
              e\in E(S)\\ n_0 \not\in e}
          }\lambda_e \sum_{j=0}^{k-1} {k \choose j+1} \E[\operatorname{Exp}(D)^{j+1}]\E[X_{S\cup e}^{k-j-1}] \label{SU4}\\
          &= \frac{k}{D}\lambda_0(S)\E[\operatorname{Exp}(D)^{k-1}] \nonumber\\
          &\quad + \frac{k}{D}\sum_{\substack{
              e\in E(S)\\ n_0 \not\in e}
          }\lambda_e \sum_{j=0}^{k-1} {k \choose j} \E[\operatorname{Exp}(D)^{j}]\E[X_{S\cup e}^{k-j-1}] \label{SU5}\\
          &= \frac{k}{D}\Biggr( \sum_{\substack{
              e\in E(S)\\ n_0 \in e}} \lambda_e \E[\operatorname{Exp}(D)^{k-1}] \nonumber\\
          &\quad+ \sum_{\substack{
              e\in E(S)\\ n_0 \not\in e}}\lambda_e \E[(X_{S\cup e} + \operatorname{Exp}(D))^{k-1}]\Biggr) \label{SU6}\\
        &= k\Biggr( \sum_{e\in E(S)} \E[X_S^{k-1}|Y=e]\P[Y=e] \Biggr) \label{SU7}\\
        &= k\E[X_S^{k-1}], \label{SU8}
    \end{align}
    where in \eqref{SU4} we expand $\E[\operatorname{Exp}(D)^k] = \frac{k!}{D^k}$ in the first term, and re-index the sum to start from $j=0$ instead of $j=1$, in \eqref{SU5} we expand $\E[\operatorname{Exp}(D)^j]$ inside the sum, pull the $k$ out of the sum, and absorb the $j+1$ in the binomial term, in \eqref{SU6} we apply the binomial theorem since the random variables are independent, in \eqref{SU7} we apply \eqref{eq:expectation_ids}, and in \eqref{SU8} we apply the law of total expectation.
\end{Proof}

\section{Conclusion and Remarks}
We have focused on reproving some identities that have a large influence on the research directions for age processes in networks, we have not made an effort to re-prove all AoI identities that exist in the literature.  Notable identities we have excluded appear in Abd-Elmagid and Dhillon~\cite{AbdElmagid2023DistributionOT}.  Those authors proved joint moment recursions of the form $\E[v_S^k v_U^j]$.  We believe that we should be able to recover these identities using similar techniques presented here, but the algebra involved is tedious and the pedagogical value of doing so is limited---we leave this as an open problem.

Our proofs crucially rely on the first passage percolation interpretation of the AoI in gossip networks.  As discussed in~\cite{maranzatto2024fpp}, Poissonian gossip networks with the standard AoI metric are very special, in that the first passage percolation formulation \textit{exactly} recovers the distribution of the age process.  Auffinger and Tang~\cite{Auffinger_revisited} managed to push the second moment analysis (a direct inspiration for our proof of the $k$th moments) to random variables which look sufficiently close to an exponential distribution near 0.  We leave as an open problem the question of whether a similar technique can be applied to Theorem~\ref{thm:moment} in gossip networks with renewal processes on edges that are sufficiently close to Poissonian.

Finally, we note that while inequality variants of \eqref{eq:yates} have appeared in the statistical physics literature, we were unable to find reference to the full recursive equality.  This is only an extra step of algebra, and the statement makes intuitive sense since the Markov property holds for $\operatorname{Exp}(1)$ random variables.  Letting $\Z^d_\ell = \{x \in Z^d : \|x\|_1 \leq \ell\}$ be the $d$-dimensional box with side length $\ell$, it is not hard to show that the \textit{time constant} for $\Z^d$ with $i.i.d.$ $\operatorname{Exp}(1)$ edge weights is,
\begin{align}
    \mu(e_1) = \liminf_{\ell \to\infty} v_{\ell, \{0\}},
\end{align}
with $v_{\ell, S}$ the recursion over the box $\Z^d_\ell$ defined as,
\begin{align}
    v_{\ell, S} = \frac{1 + \sum_{e \in E(S)} v_{\ell, S\cup e}}{|E(S)|}.
\end{align}
Then, each term in the $\lim\inf$ is in principle computable in finite time, and $\mu(e_1)$ can be approximated to arbitrary precision by taking a large enough value of $\ell$.  We attempted to use this simple observation to improve results on the time constant and Eden conjecture, but found the resulting computations intractable.  One can view the recursive equality as encoding the combinatorics that makes this classical problem notoriously difficult.

\bibliographystyle{IEEEtran}

\bibliography{bibliography}

\begin{thebibliography}{10}
\providecommand{\url}[1]{#1}
\csname url@samestyle\endcsname
\providecommand{\newblock}{\relax}
\providecommand{\bibinfo}[2]{#2}
\providecommand{\BIBentrySTDinterwordspacing}{\spaceskip=0pt\relax}
\providecommand{\BIBentryALTinterwordstretchfactor}{4}
\providecommand{\BIBentryALTinterwordspacing}{\spaceskip=\fontdimen2\font plus
\BIBentryALTinterwordstretchfactor\fontdimen3\font minus \fontdimen4\font\relax}
\providecommand{\BIBforeignlanguage}[2]{{%
\expandafter\ifx\csname l@#1\endcsname\relax
\typeout{** WARNING: IEEEtran.bst: No hyphenation pattern has been}%
\typeout{** loaded for the language `#1'. Using the pattern for}%
\typeout{** the default language instead.}%
\else
\language=\csname l@#1\endcsname
\fi
#2}}
\providecommand{\BIBdecl}{\relax}
\BIBdecl

\bibitem{aoiseminal}
S.~Kaul, R.~D. Yates, and M.~Gruteser, ``Real-time status: How often should one update?'' in \emph{2012 Proceedings IEEE INFOCOM}, 2012, pp. 2731--2735.

\bibitem{Yates_survey}
\BIBentryALTinterwordspacing
R.~D. Yates, Y.~Sun, D.~R. Brown, S.~K. Kaul, E.~Modiano, and S.~Ulukus, ``Age of information: An introduction and survey,'' \emph{IEEE Journal on Selected Areas in Communications}, vol.~39, no.~5, p. 1183–1210, May 2021. [Online]. Available: \url{http://dx.doi.org/10.1109/JSAC.2021.3065072}
\BIBentrySTDinterwordspacing

\bibitem{yates_gossip_spawc}
R.~D. Yates, ``Timely gossip,'' in \emph{2021 IEEE 22nd International Workshop on Signal Processing Advances in Wireless Communications (SPAWC)}, 2021, pp. 331--335.

\bibitem{Buyukates_2022}
\BIBentryALTinterwordspacing
B.~Buyukates, M.~Bastopcu, and S.~Ulukus, ``Version age of information in clustered gossip networks,'' \emph{IEEE Journal on Selected Areas in Information Theory}, vol.~3, no.~1, p. 85–97, Mar. 2022. [Online]. Available: \url{http://dx.doi.org/10.1109/jsait.2022.3159745}
\BIBentrySTDinterwordspacing

\bibitem{gossip_dynamic}
\BIBentryALTinterwordspacing
M.~Jelasity, A.~Montresor, and O.~Babaoglu, ``Gossip-based aggregation in large dynamic networks,'' \emph{ACM Trans. Comput. Syst.}, vol.~23, no.~3, p. 219–252, aug 2005. [Online]. Available: \url{https://doi.org/10.1145/1082469.1082470}
\BIBentrySTDinterwordspacing

\bibitem{kaswan2023age}
P.~Kaswan, P.~Mitra, A.~Srivastava, and S.~Ulukus, ``Age of information in gossip networks: A friendly introduction and literature survey,'' 2023, available online at arXiv:2312.16163.

\bibitem{Kaswan_2023}
\BIBentryALTinterwordspacing
P.~Kaswan and S.~Ulukus, ``Age of information with non-{P}oisson updates in cache-updating networks,'' in \emph{2023 IEEE International Symposium on Information Theory (ISIT)}.\hskip 1em plus 0.5em minus 0.4em\relax IEEE, Jun. 2023. [Online]. Available: \url{http://dx.doi.org/10.1109/ISIT54713.2023.10207012}
\BIBentrySTDinterwordspacing

\bibitem{maranzatto2024age}
T.~J. Maranzatto, ``Age of gossip in random and bipartite networks,'' in \emph{2024 IEEE International Symposium on Information Theory (ISIT)}, 2024, pp. 1173--1178.

\bibitem{gossip_grid}
\BIBentryALTinterwordspacing
A.~Srivastava and S.~Ulukus, ``Age of gossip on a grid,'' in \emph{2023 59th Annual Allerton Conference on Communication, Control, and Computing (Allerton)}.\hskip 1em plus 0.5em minus 0.4em\relax IEEE, Sep. 2023. [Online]. Available: \url{http://dx.doi.org/10.1109/Allerton58177.2023.10313495}
\BIBentrySTDinterwordspacing

\bibitem{yates_gossip_isit}
\BIBentryALTinterwordspacing
R.~D. Yates, ``The age of gossip in networks,'' in \emph{2021 IEEE International Symposium on Information Theory (ISIT)}.\hskip 1em plus 0.5em minus 0.4em\relax IEEE, Jul. 2021. [Online]. Available: \url{http://dx.doi.org/10.1109/ISIT45174.2021.9517796}
\BIBentrySTDinterwordspacing

\bibitem{AbdElmagid2023DistributionOT}
\BIBentryALTinterwordspacing
M.~A. Abd-Elmagid and H.~S. Dhillon, ``Distribution of the age of gossip in networks,'' \emph{Entropy}, vol.~25, 2023. [Online]. Available: \url{https://api.semanticscholar.org/CorpusID:256976733}
\BIBentrySTDinterwordspacing

\bibitem{maranzatto2024fpp}
T.~J. Maranzatto and M.~Michelen, ``Age of gossip from connective properties via first passage percolation,'' 2024, available online at arXiv:2409.12710.

\bibitem{Dhar1988FirstPP}
\BIBentryALTinterwordspacing
D.~Dhar, ``First passage percolation in many dimensions,'' \emph{Physics Letters A}, vol. 130, pp. 308--310, 1988. [Online]. Available: \url{https://api.semanticscholar.org/CorpusID:120968408}
\BIBentrySTDinterwordspacing

\bibitem{Dhar1986AsymptoticSO}
\BIBentryALTinterwordspacing
------, ``Asymptotic shape of eden clusters,'' 1986. [Online]. Available: \url{https://api.semanticscholar.org/CorpusID:118574503}
\BIBentrySTDinterwordspacing

\bibitem{Bertrand_2018}
\BIBentryALTinterwordspacing
Q.~Bertrand and J.~Pertinand, ``Dimension improvement in dhar’s refutation of the eden conjecture,'' \emph{Physics Letters A}, vol. 382, no.~11, p. 761–765, Mar. 2018. [Online]. Available: \url{http://dx.doi.org/10.1016/j.physleta.2018.01.025}
\BIBentrySTDinterwordspacing

\bibitem{Auffinger_revisited}
\BIBentryALTinterwordspacing
A.~Auffinger and S.~Tang, ``On the time constant of high dimensional first passage percolation, revisited,'' \emph{Electronic Journal of Probability}, vol.~30, no. none, Jan. 2025. [Online]. Available: \url{http://dx.doi.org/10.1214/25-ejp1274}
\BIBentrySTDinterwordspacing

\bibitem{auffinger2017fifty}
\BIBentryALTinterwordspacing
A.~Auffinger, M.~Damron, and J.~Hanson, \emph{50 years of first-passage percolation}, ser. University Lecture Series.\hskip 1em plus 0.5em minus 0.4em\relax American Mathematical Society, Providence, RI, 2017, vol.~68. [Online]. Available: \url{https://doi.org/10.1090/ulect/068}
\BIBentrySTDinterwordspacing

\end{thebibliography}
\end{document}